\documentclass[conference]{IEEEtran}

\usepackage{cite}
\usepackage{amsmath,amssymb,amsfonts}
\usepackage{amsthm}
\usepackage{algorithm}
\usepackage{algpseudocode}

\usepackage{graphicx}
\usepackage{textcomp}
\usepackage{xcolor}
\usepackage{subcaption}
\usepackage{multirow}
\usepackage{booktabs}
\newcommand{\ra}[1]{\renewcommand{\arraystretch}{#1}}
\def\BibTeX{{\rm B\kern-.05em{\sc i\kern-.025em b}\kern-.08em
    T\kern-.1667em\lower.7ex\hbox{E}\kern-.125emX}}

\begin{document}

\title{Crown Jewels Analysis using Reinforcement Learning with Attack Graphs}

\author{Rohit Gangupantulu$^{a}$,
        Tyler Cody$^{b}$$^{*}$,
        Abdul Rahman$^{b}$,\\
        Christopher Redino$^{c}$, 
        Ryan Clark$^{c}$,
        Paul Park$^{a}$\\
        \small $^{a}$Deloitte Consulting, LLC \\
        \small $^{b}$Hume Center for National Security and Technology, Virginia Polytechnic University \\
        \small $^{c}$Deloitte \& Touche LLP \\
        \small $^{*}$Corresponding author: Tyler Cody; tcody@vt.edu \\
}

\maketitle

\begin{abstract}

Cyber attacks pose existential threats to nations and enterprises. Current practice favors piece-wise analysis using threat-models in the stead of rigorous cyber terrain analysis and intelligence preparation of the battlefield. Automated penetration testing using reinforcement learning offers a new and promising approach for developing methodologies that are driven by network structure and cyber terrain, that can be later interpreted in terms of threat-models, but that are principally network-driven analyses. This paper presents a novel method for crown jewel analysis termed CJA-RL that uses reinforcement learning to identify key terrain and avenues of approach for exploiting crown jewels. In our experiment, CJA-RL identified ideal entry points, choke points, and pivots for exploiting a network with multiple crown jewels, exemplifying how CJA-RL and reinforcement learning for penetration testing generally can benefit computer network operations workflows.

\end{abstract}


\section{Introduction}

Cyber attacks pose existential, nation-level threats to electrical and financial infrastructure \cite{glenn2016cyber, varga2021cyber}, thereby directly challenging societal stability and domestic prosperity. Importantly, cyber attacks, such as the infamous Colonial Pipeline Hack \cite{reeder2021cybersecurity}, target domestic prosperity by way of enterprises \cite{varga2021cyber}. 

Enterprises and organizations generally depend on key information technology (IT) systems, known as crown jewels (CJs), operating as designed. The criticality and importance of CJs can be described in terms of the function and data they host \cite{Musman:SysEngCrownJewelsEstimation}. According to MITRE \cite{MITRE:CrownJewelsAnalysis}, ``Crown Jewels Analysis (CJA) is also an informal name for Mission-Based Critical Information Technology Asset Identification. It is a subset of broader analyses that identify all types of mission-critical assets.'' Commonly, adversaries compromise CJs not by way of ingenious technical methods, but rather by the apt use of cyber terrain \cite{conti_raymond_2018}. Adversaries often gain access to key terrain slowly over time \cite{Guion:CyberTerrainMissionMapping}.

There has been a growing interest in using notions of cyber terrain in cyber engineering practice \cite{Guion:CyberTerrainMissionMapping, applegate2017searching, conti_raymond_2018}. Chief among the reasons is an increasing appreciation for the integral roles that networks, their structure and configuration, and the paths taken in them play in engineering for cybersecurity, deterrence, and cyber resilience. While the MITRE ATT\&CK framework \cite{strom2018mitre}, for example, provides a temporally organized compendium of tactics and techniques used by adversaries, it is not clear that attack campaigns and defensive measures can be well engineered by selecting and aggregating elements from the framework. An ordered list of techniques may provide a meaningful post-hoc accounting of attack campaigns. But, for most operators, information regarding a network's structure, e.g., as captured by an attack graph, an understanding of a network's cyber terrain, and the overall path structure within the network are foundational to operational application of MITRE ATT\&CK and similar frameworks.

This paper presents a novel method for using reinforcement learning in crown jewel analysis, termed CJA-RL, that incorporates network structure, cyber terrain, and path structure to enhance operator workflow by automating the construction and analysis of attack campaigns from various entry points or initial nodes to a CJ's 2-hop network. Initial nodes and terminal nodes in the 2-hop network are analyzed in terms of footprint or quietness, i.e., the number of hops, as well as the efficient use of exploits and vulnerabilities, and, furthermore, paths are analyzed for pivot and choke points \cite{conti_raymond_2018}.

In the cybersecurity literature, although cyber terrain has been advocated for \cite{Guion:CyberTerrainMissionMapping, applegate2017searching ,conti_raymond_2018, larkin2021towards}, it has not been widely applied to automated methods like machine learning. Gangupantulu \textit{et al.} present methods for building cyber terrain into Markov Decision Process (MDP) models of attack graphs \cite{gangupantulu2021using}. Specifically, firewalls are treated as cyber obstacles \footnote{Obstacles are part of the OAKOC intelligence preparation of the battlefield framework described in \cite{conti_raymond_2018}}. In contrast, this paper contributes a method for identifying key terrain and avenues of approach by probing networks with RL. As such, CJA-RL introduces cyber terrain to penetration testing by way of the specific use of RL, as opposed to augmenting state dynamics or rewards in MDP models of attack graphs. Since methods of modeling attack graphs and MDPs can vary between networks and enterprises, this CJA-RL approach provides a more general mechanism for incorporating terrain, albeit different terrain, than prior work \cite{gangupantulu2021using}.

The paper is structured as follows. First, a background is given on crown jewels, reinforcement learning, and reinforcement learning for penetration testing. Then, the CJA-RL method is presented and results are discussed. The paper concludes with a synopsis and remarks on how CJA-RL can augment cybersecurity practices.

\section{Background}

\subsection{Crown Jewels}

Typically, mission assurance and resiliency processes start with CJA as a methodology to align an organization's mission with their critical cyber assets \cite{MITRE:CrownJewelsAnalysis}. Prior to initiating such an analysis, the critical needs of the business mission must be defined. These findings can be employed to scope and size the infrastructure required to fulfill the mission need.  The IT assets that align to this mission are considered to be the most important and of the highest value; they are the ``Crown Jewels'' of the organization.

The assignment of a mission relevance value for these IT assets lies at the heart of CJA \cite{MITRE:CrownJewelsAnalysis}. The depth of such an analysis depends on the functional use of the IT asset by the organization. For example, a payroll system may be a target as it has not only information related to salaries but highly sought after personal identifying information (PII) in a single location. Such data, once exfiltrated, can be sold at a high price or used for nefarious purposes. By measuring the importance of these IT assets, the organization is capable of assigning risk and value scores for each, and CJs can be determined computationally \cite{Guion:CyberTerrainMissionMapping}. 

In real-world cases of penetration testing and adversary emulation, attackers do not always follow direct or expected attack paths \cite{weissman1995penetration, chen2018penetration}. The assumption that attackers will follow direct paths when performing CJA creates gaps in assigned risks and values. While MITRE's ATT\&CK Framework highlights methodologies used to exploit these gaps \cite{strom2018mitre}, organizations struggle to derive the secondary or tertiary effects of these exploits. As a result, such gaps create a situation where reconnaissance and access exploits during late-stage attacks can appear to have little impact on an organization's CJs, when in reality adversaries are using them to position themselves near key terrain and avenues of approach to further their attack campaign goals. Consider the following example of such a phenomena.

\subsubsection{Motivating Example}


A common cause of gaps between true and assigned risks and values is the mis-evaluation of the risk associated with visibility and network exposure. A real-world example of this is a print server. In evaluating a print server in CJA, an organization may make the assumption that printing services being inoperable or degraded will have no immediate impact on the organization's business, nor impact other critical IT assets. However, the print server could serve as a pivot point\footnote{Pivot points are footholds in networks for lateral movement.} within the network as it may lie within a 1- or 2-hop graph of a target CJ. By inappropriately identifying the risk and value of a print server within close proximity to a CJ, the organization will underestimate its value in an adversarial attack campaign. This failure results from ignoring the need for identifying exploitable terrain (i.e. print servers and other infrastructure) within the breadth of the enterprise integrated services in complex infrastructures \cite{geer2002penetration, denis2016penetration}. In addition to printing, users in the midst of their day-to-day tasks depend on services like authentication through identity and access management (IdAM), network access to key business shares, and connectivity to the internet from within the enterprise \cite{conti_raymond_2018}. The integration of such services forms tertiary yet key aspects in the analysis of CJ that are critical to business function. The method in this paper, CJA-RL, offers a means of decreasing these tangential attack opportunities by identifying key terrain and avenues of approach and by identifying risks based on network structure and cyber terrain; contrary to most gap-prone threat-model-driven assessments.


\subsection{Reinforcement Learning}

RL concerns agents who learn by taking actions in an environment and receiving rewards \cite{sutton2018reinforcement}. An agent interacts with an environment $\mathcal{E}$ by taking actions $a_t$ and receiving states $s_{t+1}$ and rewards $r_{t+1}$. Typically an MDP is used to model $\mathcal{E}$. A finite MDP is a tuple $\langle S, A, \Phi, P, R \rangle$, where $S$ is a set of states, $A$ is a set of actions, $\Phi \subset S \times A$ is the set of admissible state-action pairs, $P:\Phi \times S \to [0, 1]$ is the transition probability function, and $R: \Phi \to \mathbb{R}$ is the expected reward function where $\mathbb{R}$ is the set of real numbers. $P(s_t, a_t, s_{t+1})$ denotes the transition probability from state $s_t$ to state $s_{t+1}$ under action $a_t$, and $R(s, a)$ denotes the expected reward from taking action $a$ in state $s$.  

RL agents learn by maximizing expected future rewards. Specifically, the discounted sum of future rewards is 
\begin{equation} \label{discounted_sum}
    R_t = \sum_{k=0}^\infty \gamma^k r_{t+k}
\end{equation}
the action value function is 
\begin{equation} \label{action_value}
    Q^\pi(s, a) = \mathbb{E}[R_t|s_t=s, a]
\end{equation}
and the optimal action-value function is
\begin{equation} \label{optimal_action_value}
    Q^*(s, a) = \max_\pi Q^\pi (s, a)
\end{equation}
where $\gamma \in (0, 1)$ is a discount factor and $\pi$ is a policy mapping states and actions $(s, a)$ to the probability of picking action $a$ in state $s$.

In this paper, Deep Q-learning (DQN) is used to approximate $Q^*$ with a neural network $Q(s, a; \theta)$, where $\theta$ are parameters of the neural network \cite{mnih2013playing, mnih2015human}. DQN has been widely used in the penetration testing literature \cite{schwartz2019autonomous, chowdary2020autonomous, hu2020automated, gangupantulu2021using}.

\subsection{Reinforcement Learning for Penetration Testing}

RL has seen recent interest by the cybersecurity community for its variety of benefits. Penetration testing is multi-faceted \cite{austin2011one}, and RL is general enough to be the basis for many tools for the various types of penetration testing, e.g., external testing, internal testing, blind testing, and double-blind testing \cite{weissman1995security}. Additionally, RL solution methods can scale penetration testing to the challenges of future networks, such as payload mutation and intelligent entry-point crawling \cite{chen2018penetration}.

\begin{figure*}[t]
    \centering
    \includegraphics[width=\textwidth]{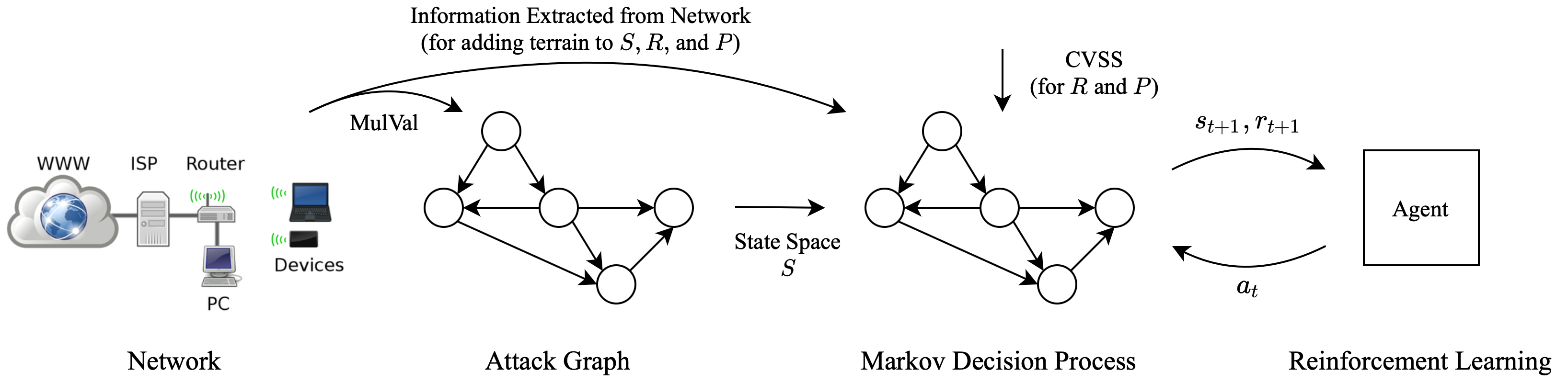}
    \caption{The Process of Deploying RL on MDPs Modeled using Attack Graph Representations of Networks.}
    \label{fig:pentestingRL}
\end{figure*}

So far, RL methods for penetration testing use the attack graph model \cite{ghanem2018reinforcement, schwartz2019autonomous, ghanem2020reinforcement, chaudhary2020automated, yousefi2018reinforcement, chowdary2020autonomous, hu2020automated, gangupantulu2021using}. Figure \ref{fig:pentestingRL} depicts the high-level process of deploying RL on a network. Most authors treat $\mathcal{E}$ as an MDP, as is done herein, where network structure and configuration are observable. While treating $\mathcal{E}$ as a partially observable MDP (POMDP), where structure and configuration are uncertain, is more realistic, such approaches have not been shown to scale to large networks and require modeling many prior probability distributions \cite{shmaryahu2016constructing}. RL agents are trained in an episodic fashion where terminal states are the targets of penetration testing.

Similar to previous literature, vulnerability information is used to construct the MDP \cite{yousefi2018reinforcement, hu2020automated, chowdary2020autonomous, gangupantulu2021using}. More specifically, following Gangupantulu \textit{et al.} \cite{gangupantulu2021using}, the MDPs used in this paper incorporate notions of cyber terrain by treating firewalls as cyber obstacles by lowering corresponding transition probabilities in MDPs. In contrast to prior work, a larger network is used herein relative to previous literature on RL for penetration testing, as reported in Table \ref{table:size}.

\begin{table*}[t]
\centering
\ra{1.3}
\begin{tabular}{@{}ll@{}}
\toprule
Paper & Network Description(s) \\
\midrule
Ghanem and Chen \cite{ghanem2018reinforcement} & \emph{100 machine} local area network \\
Schwartz and Kurniawati \cite{schwartz2019autonomous} & \emph{50 machines} with unknown services and 18 machines with 50 services \\
Ghanem and Chen \cite{ghanem2020reinforcement} & \emph{100 machine} local area network \\
Chaudhary \textit{et al.} \cite{chaudhary2020automated} & Not reported \\
Yousefi \textit{et al.} \cite{yousefi2018reinforcement} & Attack graph with \emph{44 vertices and 43 edges} \\
Chowdary \textit{et al.} \cite{chowdary2020autonomous} & Attack graph with 109 vertices, edges unknown, and a \emph{300 host} flat network \\
Hu \textit{et al.} \cite{hu2020automated} & Attack graph with \emph{44 vertices and 52 edges} \\
Gangupantulu \textit{et al.} \cite{gangupantulu2021using} & Attack graph with \emph{955 vertices and 2350 edges} \\
\textbf{Our largest network} & Attack graph with \textbf{1617 vertices and 4331 edges} \\
\bottomrule
\\
\end{tabular}
\caption{The largest network sizes in the RL penetration testing literature as reported in \cite{gangupantulu2021using}.}
\label{table:size}
\end{table*}

The attack graph is constructed by passing network information to MulVal \cite{mulvalouarticle}. MulVal uses a reasoning engine to conduct vulnerability analysis and generate a graphical network representation. In modeling MDPs using attack graphs, the state space $S$ consists of the attack graph's vertices and the actions $A$ consist of its edges. The vertices can be components of the network or means of traversal, e.g., intermediary file servers or interaction rules, respectively, and connections between components and means of traversal are the actions $A$.

The Common Vulnerability Scoring System (CVSS) \cite{mell2006common} is used to model the transition probabilities $P$ and rewards $R$. The CVSS is an open and industry-standard method of measuring cybersecurity vulnerabilities by using a database of known exploits. The CVSS offers an automated approach to constructing attack graphs and MDPs for RL that is common in the literature \cite{yousefi2018reinforcement, chowdary2020autonomous, hu2020automated, gangupantulu2021using}. In this paper, following Hu \textit{et al.} and Gangupantulu \textit{et al.} \cite{hu2020automated, gangupantulu2021using}, the transition probabilities $P(s_t, a_t, s_{t+1})$ are modeled as $0.9$, $0.6$, and $0.3$ of transitioning out of $s_t$ when the attack complexity of $s_{t+1}$ is low, medium, and high, respectively. The reward $r_{t+1}$ is given by
\begin{equation}
    \emph{Base Score}(s_{t+1}) + \frac{\emph{Exploitability Score}(s_{t+1})}{10}
\end{equation}
when the agent successfully transitions and $0$ otherwise. In addition to this reward, reward of 100 is given for arriving in the terminal node and a reward linearly scaled from 0 to 100 is given for arriving at each node along the depth first search path from the initial to terminal node. Lastly, cyber terrain is including in the MDP via treating firewalls as cyber obstacles. Following Gangupantulu \textit{et al.}, firewall information is included in the modeling of the MDP by downwardly adjusting transition probabilities corresponding to whether FTP, SMTP, HTTP, or SSH is used \cite{gangupantulu2021using}.

\section{Methods}

Performing CJA using RL involves multiple steps of domain understanding and graph analysis. The attack graph extracted using MulVal is parsed multiple ways. Firstly, following \cite{gangupantulu2021using}, the input.p file to the attack graph generator as well as the resultant set of attack graphs is examined. The input.p file contains valuable information about the nature of the network, such as subnets and their connections, firewalls, and sensitive hosts. The attack graph is a snapshot of the network oriented towards a representing the capabilities available to a penetration tester, such as privilege escalation opportunities, the range of exploits available, and the level of vulnerabilities that lie in a network. Using domain understanding, CJs are found by assessing their importance to the system and its operational ability as a whole. In this work, these CJs were either Linux-based objects within a system, such as a wireless server or the root user account, or Active Directory objects, such as a specific machine in an organization's network infrastructure.

Upon identifying the CJ within the attack graph, the next steps are to construct the network's MDP and to identify the CJ's 2-hop network. Then, the set of initial nodes wherefrom the CJ's 2-hop network is reachable is identified. This involves iteratively assessing possible paths to the 2-hop network. Once a set of reachable initial nodes is established, the CJA-RL algorithm can be run.

\begin{figure}[t]
    \centering
    \includegraphics[width=0.5\textwidth]{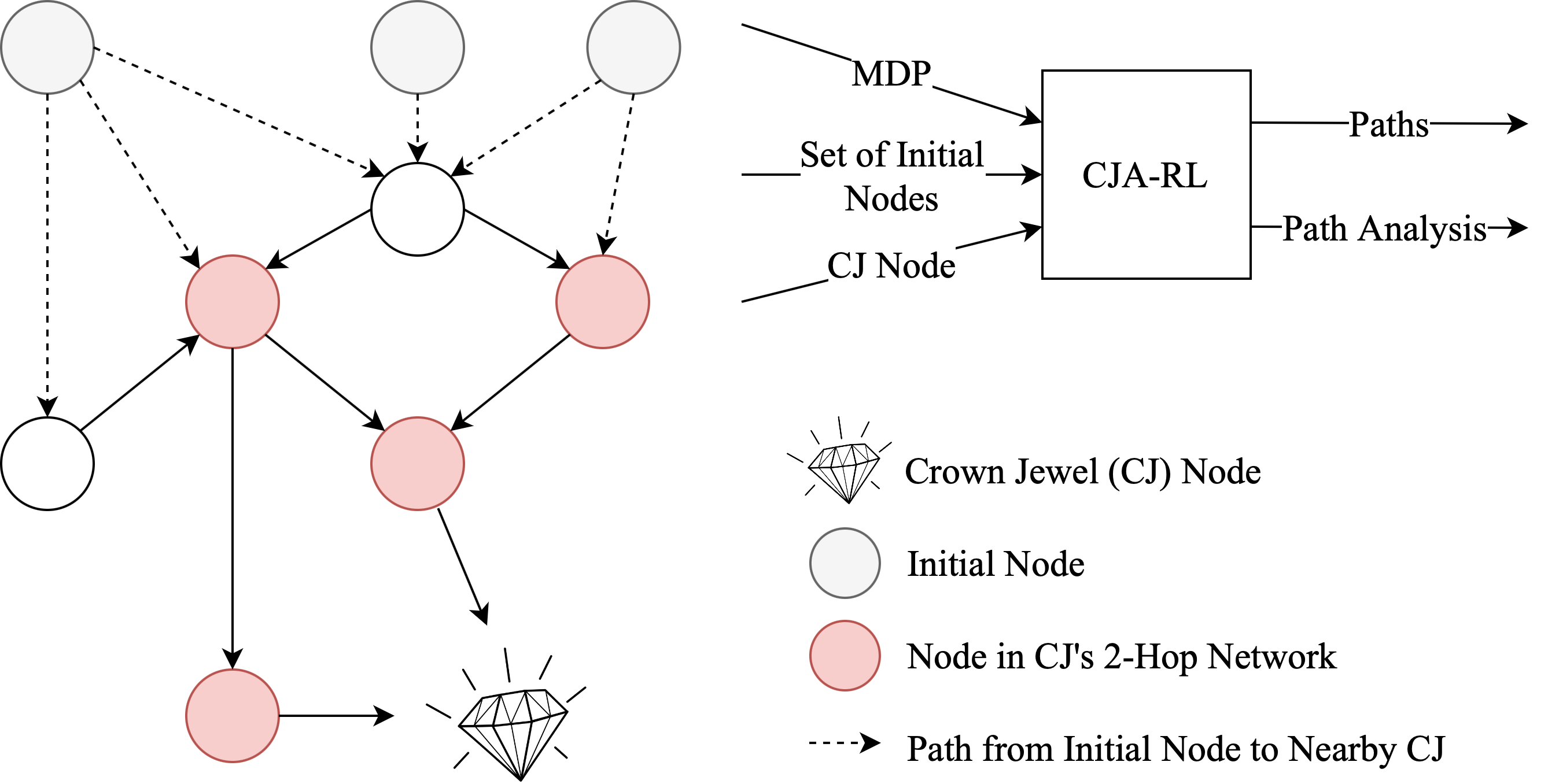}
    \caption{Depiction of CJA-RL.}
    \label{fig:cjarl}
\end{figure}

CJA-RL takes the MDP, CJ node, and reachable initial nodes as inputs. The CJ's 2-hop network is found, then for each initial node and each node in the CJ's 2-hop network, an RL agent is episodically trained until total reward converges. The optimal learned policy, i.e., $Q^*$ in Equation \ref{optimal_action_value}, is used to find the optimal path. The paths are collected and then analyzed using statistics and manual inspection. Algorithm \ref{alg:cjarl} follows this description and Figure \ref{fig:cjarl} provides depictions of CJA-RL.

High reward corresponds to efficient use of exploits and low number of hops corresponds to being low-footprint or quiet. These metrics can be used to analyze key terrain by ranking initial nodes, i.e., network entry points, and terminal nodes, i.e., pivots to the CJ. Also, nodes frequently while traversing the network can be used to characterize avenues of approach, e.g., choke points. The outcome of CJA-RL is cyber-terrain-oriented understanding of the hosts in a network that are advantageous or otherwise useful for a malicious actor to compromise in order to gain access to CJs. 

\begin{algorithm}[t]
\caption{Crown Jewel Analysis via RL (CJA-RL)}\label{alg:cap}
\begin{algorithmic}
\Require MDP, CJ node, reachable initial nodes
\Ensure Paths to CJ's 2-hop network and analysis

\State $\emph{2-hop} \gets f_{\emph{2-hop}}(CJ)$ \Comment{Get nodes in 2-hop network}

\For{$i$ in reachable initial nodes}
    \For{$j$ in \emph{2-hop}}
        \State $path \gets f_{RL}(MDP, i, j)$ \Comment{Optimal path $i \to j$}
        \State $paths \gets store(path)$
\EndFor
\EndFor \\
\Return $paths$, $analysis(paths)$
\label{alg:cjarl}
\end{algorithmic}
\end{algorithm}

\section{Results}

Our formalism for CJ path determination illuminates those avenues of approach for an actor that hopes to access inconspicuous hosts on the road to exploiting key IT assets while remaining undetected. Hosts that lie within the 2-hop graph of a CJ, that possess vulnerabilities, applications, or elements that can be exploited, offer real value in adversarial attack campaigns. Gaining footholds within proximity of a CJ is often a needed intermediate step for launching additional capabilities needed to fully exfiltrate or compromise a CJ \cite{conti_raymond_2018}.

When assessing the performance of the approaches that were taken, it is prudent to note that this is not a one-dimensional assessment of rewards accrued by the agent - as that would fail to paint clear context behind the CJ and its neighborhood. Instead, CJA-RL is applied to a few examples of different CJs that have been identified in a network. The results assess both the nature of the network itself that is being utilized, as well as the validity of CJA-RL and how it can help drive realistic outcomes during the exercise of penetration testing.

Analysis identified the best terminal nodes, most visited nodes, and best initial nodes for a set of CJs. The terminal node that has the least number of hops of traversal needed to reach that node, while maximizing the reward, has the benefit of being the quietest staging node in which to gain a foothold. It also ends up becoming the ideal candidate from which to stage an attack campaign on a CJ. This kind of analysis introduces the concept of cover and concealment, found often in cyber terrain analysis \cite{conti_raymond_2018}. The most visited nodes has the importance of being ideal candidates for sensors/trackers. It is also important to consider the proportion of paths that utilize the most visited node in designing sensor/tracker systems. The statistics regarding RL performance from the different initial nodes can be used to identify which entry points are best for which CJs. Additionally, initial nodes can be studied on an aggregate basis to assess if there lies a single initial node that is best for attacking multiple CJs. This is particularly useful information for penetration testers looking to launch attack campaigns which pivot laterally across the enterprise.

\begin{table*}
\resizebox{\textwidth}{!}{%
\begin{tabular}{@{}llll@{}}
\toprule
\textbf{Crown Jewel} &
  \textbf{Best Initial Node} &
  \textbf{Best Terminal Node} &
  \textbf{Most Visited Node} \\ \midrule
{AD Computer - comp00384} &
  \multicolumn{1}{|l|}{attack on abraniff00503 in comp00384} &
  ruleHasActiveSession(cclinker00753) &
  \multicolumn{1}{|l}{RULE 1 (Exploit active session)} \\ \cmidrule(r){1-4}
{AD Computer - fllabdc} &
  \multicolumn{1}{|l|}{attack on lcopa00557 in fllabdc} &
  attack on misby00484 in fllabdc &
  \multicolumn{1}{|l}{RULE 1 (Exploit active session)} \\ \cmidrule(r){1-4}
{AD Group - administrators} &
  \multicolumn{1}{|l|}{Active Session exploited  for lrainer00755} &
  attack on sfordham00591 in comp00001 &
  \multicolumn{1}{|l} {RULE 28 (Users with active login sessions)} \\ \cmidrule(r){1-4}
{Object in subnet network\_core} &
  \multicolumn{1}{|l|}{attack on machine dbranstad00508} &
  \multicolumn{1}{|l|} {root access to cloud\_conn\_mgmt} &
  \multicolumn{1}{|l} {RULE 1 (Exploit active session)} \\ \cmidrule(r){1-4}
 {mail\_server object in any subnet} &
  \multicolumn{1}{|l|}{attack on efeltus00249 in comp00955} &
  \multicolumn{1}{|l|} {attack on wfeagins00889} &
  \multicolumn{1}{|l} {RULE 28 (Users with active login sessions)} \\ \bottomrule
\end{tabular}%
}
\caption{The table shows the statistics behind the RL execution for a variety of crown jewels, indicating their best initial node, best terminal node, and most visited node. Here, the `best' nodes are determined by minimizing the number of hops to reach the crown jewel neighborhood while maximizing the total reward.}
\label{tab:table2results}
\end{table*}



Table \ref{tab:table2results} give a snapshot of CJA-RL applied to multiple CJs in our network. It appears that the most visited node tends to be an interaction rule of either an active session or an active login session. These are easily exploitable means of navigating the network across subnets to enter the CJ's neighborhood. Both network defenders and adversaries could consider this type of host as an excellent candidate to pivot laterally across the network, thus driving subsequent methods for exploitation or defense. Additionally, attacks on various host computers within a subnet tend to be the best terminal nodes. This tells us that within an attack graph setting, engaging via interaction rules is the way of preferred traversal and the gateway into a specific CJ target. This illuminates some of the subtle methods used by attackers to traverse networks while not triggering deployed countermeasures. Awareness of where these hosts exist are critical for suggesting the path an adversary could use to compromise a CJ or how defenders can deploy defensive countermeasures for CJ protection.

Additionally, following the OAKOC cyber terrain principles of key terrain and avenues of approach \cite{conti_raymond_2018}, by looking at what nodes in the CJ's neighborhood require the least number of hops to get to. In our case, these nodes where an augmentation of a Bloodhound-based Active Directory domain \cite{bloodhound}, through a relationship established between nodes in the attack graph.

\section{Conclusion}

This paper presents methods enhancing previous use of cyber terrain in reinforcement learning for penetration testing \cite{gangupantulu2021using} by using the concept of a crown jewel to identify advantageous hosts within a network. Our methodology maintains an automated, scale-oriented approach to constructing MDPs, while introducing notions of cyber terrain and path analysis that help users understand their networks on a deeper level. 

CJA-RL is a kind of cyber terrain analysis and is based on cyber terrain methodologies within intelligence preparation of the battlefield. We showed how CJA-RL can be used to find some of the most critical hosts within a network, including: ideal entry points, ideal pivots for sensors/trackers, ideal footholds nearby crown jewels, and hosts that could be used to compromise an entire system, i.e., multiple CJs, if penetrated.

CJA-RL's rich, empirical, and automated output allows operators and engineers to quickly develop an understanding of network and terrain, allowing them to focus their time and expert knowledge elsewhere. Two examples of this are (1) building richer attack campaigns that couple threat models with the terrain and paths to CJs identified by CJA-RL or (2) instituting defensive measures near key terrain and related avenues of approach. 

Future work should consider how more elements of cyber terrain can be folded into MDP construction to increase quality of advantageous host predictions. One approach is to use a labeling procedure. A labeling procedure would have the added benefit of producing further context for path recommendations within a neighborhood. However, if labeling is introduced, it should carefully consider how it limits the scale the size of attack graphs. CJA-RL can be applied to richer attack graphs and MDPs, and such an extension would further validate CJA-RL as well as the use of intelligence preparation of the battlefield principles in creating realistic contexts for penetration testing. Methods should be developed that quantify the impact of a crown jewel in a more meaningful way than the CVSS-based reward paradigm. In addition, CJA-RL can be integrated with security incident and event management (SIEM) systems and/or security-focused data lakes as a means to recommending weaknesses within the network that are candidates for defense or countermeasure deployments.

\section{Acknowledgements}

This work was made possible through the collaboration between Dr. Michael Ambroso, Lead for the AI/ML for Cyber Testing Innovation Pipeline group---funded by Deloitte's Cyber Strategic Growth Offering led by Deborah Golden, and Dr. Laura Freeman, Director of the Hume Center for National Security and Technology's Intelligent Systems Lab at the Virginia Polytechnic Institute and State University.

\bibliographystyle{IEEEtran}
\bibliography{ref}

\end{document}